\documentclass[aps,pra,twocolumn,showpacs,preprintnumbers,amsmath,amssymb]{revtex4-1}
\usepackage{graphicx}
\usepackage{dcolumn}
\usepackage{bm}
\usepackage{enumerate}
\usepackage{color}

\begin{document}
\preprint{APS}
\title{Fluctuation properties of laser light after interaction with an atomic system: comparison between two-level and multilevel atomic transitions}
\author{A. Lezama$^{1}$, R. Rebhi$^{2,3}$, A. Kastberg$^{4}$, S. Tanzilli$^{4}$ and R. Kaiser$^{2}$}

\affiliation{$^{1}$ Instituto de F\'{i}sica, Universidad de la Rep\'{u}blica. Casilla de correo 30, 11000, Montevideo, Uruguay}
\affiliation{$^{2}$Universit\'{e} Nice Sophia Antipolis, Institut Non-Lin\'{e}aire de Nice, CNRS UMR 7335, F-06560, Valbonne, France}
\affiliation{$^{3}$Centre for Quantum technologies, National University of Singapore, Singapore}
\affiliation{$^{4}$ Universit\'{e} Nice Sophia Antipolis, Laboratoire de Physique de la Mati\`{e}re Condens\'{e}e, CNRS UMR 7336, Parc Valrose, F-06108 Nice Cedex 2, France.}
\date{\today}

\begin{abstract}
The complex internal atomic structure involved in radiative transitions has an effect on the spectrum of fluctuations (noise) of the transmitted light. A degenerate transition has different properties in this respect than a pure two-level transition. We investigate these variations by studying a certain transition between two degenerate atomic levels for different choices of the polarization state of the driving laser. For circular polarization, corresponding to the textbook two-level atom case, the optical spectrum shows the characteristic Mollow triplet for strong laser drive, while the corresponding noise spectrum exhibits squeezing in some frequency ranges. For a linearly polarized drive, corresponding to the case of a multilevel system, additional features appear in both optical and noise spectra. These differences are more pronounced in the regime of a weakly driven transition: whereas the two-level case essentially exhibits elastic scattering, the multilevel case has extra noise terms related to spontaneous Raman transitions. We also discuss the possibility to experimentally observe these predicted differences for the commonly encountered case where the laser drive has excess noise in its phase quadrature. 
\end{abstract}

\maketitle
\section{\label{introduccion}Introduction}

Light matter interaction has been at the center of intensive research for many decades. In particular, light interacting with dilute gazes of atoms has allowed to investigate a number of fundamental questions, ranging from quantum optics to mesoscopic physics or condensed matter physics.
For a precise study of light matter interaction, the internal Zeeman structure needs specific attention and can lead to qualitative differences between alkali atoms and alkali earth metal atoms. This fundamental difference has for instance allowed for the surprisingly efficient sub-Doppler cooling schemes only possible in presence of a Zeeman degenercay in the ground state \cite{Chu1998, CohenTannoudji1998, Philipps1998} or to the efficient production of nonclassical states of radiation \cite{Squeezing,LAMBRECHT96,Ries03} 

In the context of mesoscopic physics, the non degenerate Zeeman structure of the ground state has been studied in the context of coherent backscattering, where a novel ``dephasing" mechanism has been attributed to the Zeeman structure of the atoms \cite{Labeyrie2003}. Motivated by a theoretical debate whether the internal structure of atoms (Zeeman degeneracy) is source of additional noise for light transmitted through an atomic sample~\cite{Akkermans2007,Gremaud2008,Akkermans2008, Miniatura2015}, we turned to a microscopic, ab initio model, to investigate the fundamental role of the Zeeman structure on the fluctuations of propagating electromagnetic radiation after interaction with a dilute gaz of atoms. At first glance, the prediction by ~\cite{Gremaud2008, Miniatura2015} might come as a surprise, as this work predicts no impact of the internal structure of the atoms on the noise correlation function. This seems in contrast to the results for the average intensities, where the role of the internal structure is manifest and well understood. However, an ab initio microscopic model fo the correlation functions in the multiple scattering limit discussed in~\cite{Akkermans2007,Gremaud2008,Akkermans2008, Miniatura2015} is still out of range. We therefore focus on a different regime, where the ab initio model can be used and exploited for a direct qualitative and quantitative comparison between a two level system and a multilevel configuration. 
We stress that the noise spectrum as studied in this paper, despite its apparent qualitative ressemblence to the correlation function studied in ~\cite{Akkermans2007,Gremaud2008,Akkermans2008, Miniatura2015} does not rely on the same combination of operators and that we are also restricted to a low optical thickness, where multiple scattering can be neglected. Our model however allows to characterize the fluctuations of the electromagnetic radiation after interacting with an ensemble of cold atoms at a precision of the quantum level, whereas the correlation functions studied in ~\cite{Akkermans2007,Gremaud2008,Akkermans2008, Miniatura2015}  were based on classical fluctuations.
Even though the situation addressed in this work does not exactly match the configurations discussed in Refs.~\cite{Akkermans2007,Gremaud2008,Akkermans2008}, it can however readily be implemented and exploited experimentally. In particular, this could be achieved by using laser cooled atomic vapors and isolating a single transition, therefore making it possible to neglect Doppler broadening.

The effect of atomic state degeneracy on light-atom interaction has also been studied in different contexts. The manifestations that are perhaps most obvious to investigate are resonance fluorescence and the spectrum of transmitted light. In \cite{Polder1976}, the spectrum of resonance fluorescence for a $j_g=1/2$ to $j_e=1/2$ transition was calculated. Further complexity was added in later work, such as the effect on the spectrum of the presence of degeneracy and collisions~\cite{Cooper1980}, and higher degrees of degeneracy, corresponding to a concrete atom~\cite{Javanainen1992}. In the 90s, Bo Gao made thorough analytical studies of the effect of degeneracy, for an atom interacting with near resonant light, on the probe spectra and resonance fluorescence~\cite{Gao1993,Gao1994}. The effect on lineshapes was studied in \cite{Galbraith1982}.  Note that most of the above mentioned theoretical works are covered in a review article~\cite{Berman2008}.

There are fewer reports on experimental studies. In \cite{Bergen1988}, the Mollow triplet in the presence of atomic degeneracy was studied in a cell with a buffer gas. The absorption spectra of degenerate two-level atomic transitions was experimentally explored in \cite{Lezama1999}. Intensity correlations in scattered light was investigated in \cite{Walker1996}.  Squeezing of the transmitted light was studied, theoretically and experimentally, in \cite{LEZAMA08,Mikhailov09,Barreiro2011}, coherent backscattering in \cite{Labeyrie1999}, random lasing in \cite{baudouin2013cold,Guerin2009}, and a review on coherent transport phenomena in cold atoms can be found in \cite{Labeyrie2008}.\\

An alternative approach to the one we have adopted here, would be the consideration of the interaction of a Zeeman degenerate atomic system with the field contained in an optical cavity. Such approach has been extensively adopted for the study of cold atoms in  cavity QED experiments \cite{LAMBRECHT96,VERNAC02}. Particularly relevant to the problem that we are addressing here, is the generation of polarization squeezing inside an optical cavity \cite{JOSSE03} for which the Zeeman sublevel structure of the atomic transition plays an essential role. 
\section{\label{modelo}Model}

We consider here both two-level and multi level configurations as outlined in \figurename{~\ref{levels}}.

Following \cite{FABRE97}, we consider a single spatial mode field of frequency $\omega_L$ and wave-number $k$ propagating along the $z$-axis. In the Heisenberg picture, the field is described by the operator:

\begin{figure}
\includegraphics[width=8.5cm]{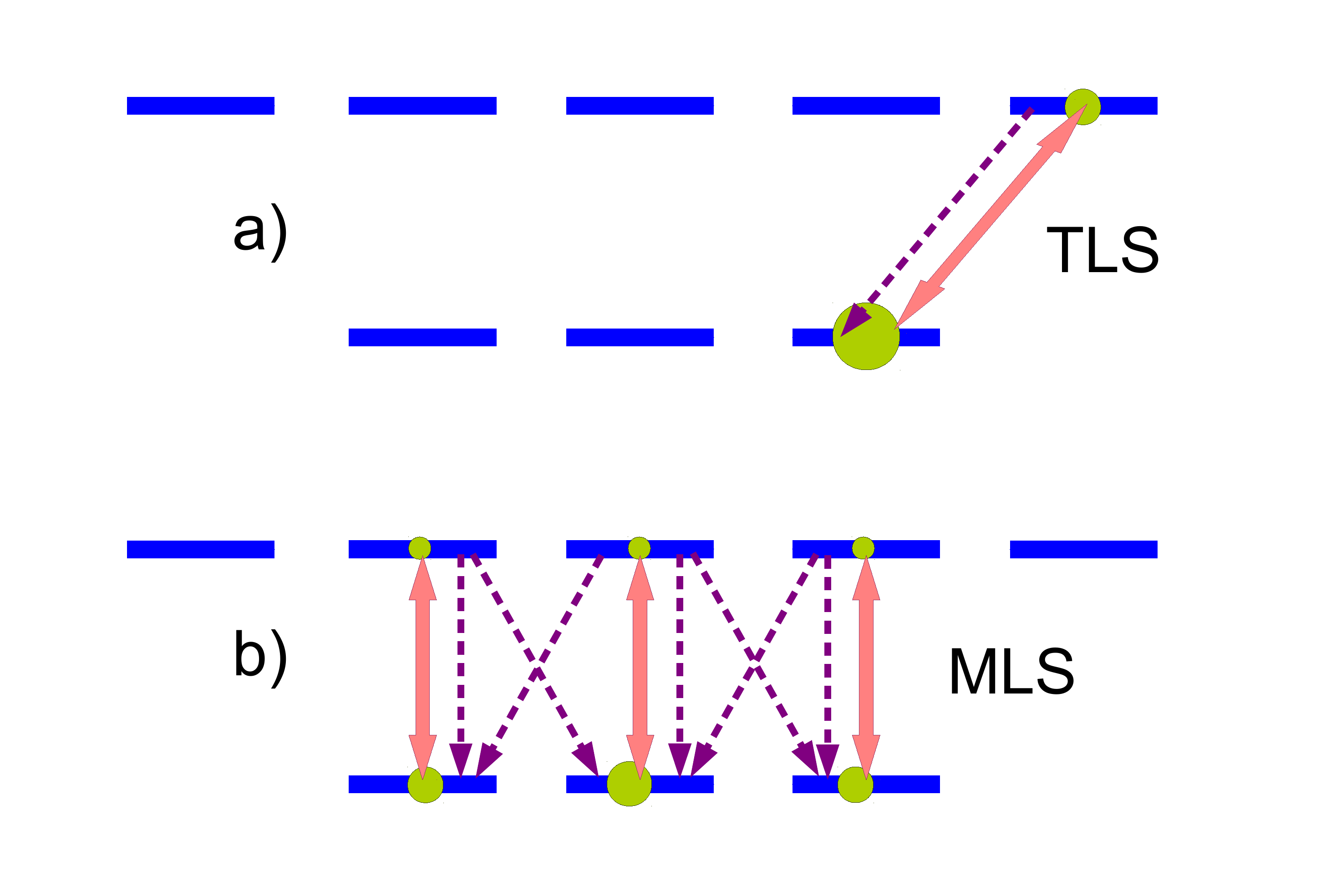}
\caption{\label{levels}(Color online) Sketch of the level configurations considered for an $F_g=1\rightarrow F_e=2$ dipolar atomic transition. a) Two-level system (TLS), circular polarization, quantization axis along light wavevector. b) Multi-level system (MLS), linear polarization, quantization axis along light polarization. Thick arrows: incident field coupling. Dashed arrows: spontaneous emission channels. Circles represent steady-state populations.}
\end{figure}

\begin{eqnarray}
\vec{E} (z,t) &=&\xi \left( a_1e^{i(kz-\omega _Lt)}%
\hat{e}_1^{*}+a_2e^{i(kz-\omega _Lt)}\hat{e}_2^{*}\right.
\label{campo} \\
&&\left. +a_1^{\dagger }e^{-i (kz-\omega _Lt) }\hat{e}%
_1+a_2^{\dagger }e^{-i (kz-\omega _Lt) }\hat{e}_2 \right)\;,
\nonumber
\end{eqnarray}
where $\xi =\sqrt{\frac{\hbar \omega_{L}}{2\epsilon_{0} AL}}$ is the single photon field amplitude, $L$ the field quantization volume length, $A$ the mode cross-section, and $\epsilon_{0}$ the vacuum permittivity. $\hat{e}_1$ and $\hat{e}_2$ are two orthogonal (complex) polarization unit vectors and $a_1$, $a_2$, $a_1^{\dagger }$, $a_2^{\dagger }$ are the slowly varying field annihilation and creation operators, obeying the commutation rules:

\begin{subequations}\label{commut}
\begin{eqnarray}
\left[ a_\kappa (z,t),a_\lambda (z^{\prime },t^{\prime }) \right] &=&0\\
\left[a_\kappa (z,t) ,a_\lambda^{\dagger } (z^{\prime },t^{\prime })
\right] &=&\delta _{\kappa \lambda}\frac Lc\delta (t-t^{\prime
}-\frac{z-z^{\prime }}c)\\
(\kappa,\lambda=1,2),\nonumber
\end{eqnarray}
\end{subequations}
with $c$ the speed of light in vacuum.\\

The field fluctuations operators are:
\begin{equation}\label{fluct}
\delta a_\lambda\equiv a_\lambda - \langle a_\lambda \rangle\qquad (\lambda=1,2),
\end{equation}
and the spectral correlation matrix $\mathcal{S}=\left\lbrace \mathcal{S}_{ij}\right\rbrace $ ($i,j=1,2$) for a given polarization component of a stationary field can be written as:
\begin{equation}
\mathcal{S}(\Omega)=\xi^{2 }\int e^{i\Omega\tau} \left( \begin{matrix}
\langle a(\tau) a^{\dagger}(0)\rangle & \langle a(\tau) a(0)\rangle\\
\langle a^{\dagger}(\tau) a^{\dagger}(0)\rangle & \langle  a^{\dagger}(\tau) a(0)\rangle
\end{matrix} \right) d\tau.
\end{equation}

Using (\ref{fluct}) the spectral correlation matrix can be separated into two terms:
\begin{align}
&\mathcal{S}= S^{E}(\Omega)+S^{IN}(\Omega)\\
&S^{E}(\Omega)\equiv\xi^{2 }\int e^{i\Omega\tau} \left( \begin{matrix}
\langle a(\tau)\rangle\langle a^{\dagger}(0)\rangle & \langle a(\tau)\rangle\langle a(0)\rangle\\
\langle a^{\dagger}(\tau)\rangle\langle a^{\dagger}(0)\rangle & \langle a^{\dagger}(\tau)\rangle\langle a(0)\rangle
\end{matrix} \right) d\tau\\
&S^{IN}(\Omega)\equiv\xi^{2 }\int e^{i\Omega\tau} \left( \begin{matrix}
\langle \delta a(\tau)\delta a^{\dagger}(0)\rangle & \langle \delta a(\tau)\delta a(0)\rangle\\
\langle \delta a^{\dagger}(\tau)\delta a^{\dagger}(0)\rangle & \langle \delta a^{\dagger}(\tau)\delta a(0)\rangle
\end{matrix} \right) d\tau.
\end{align}
Here $S^{E}(\Omega)$ describes the \textit{elastic} (classical) fluctuations associated with the variations of the field mean value. $S^{IN}(\Omega)$ describes the inelastic contribution to light fluctuations, which have a quantum mechanical origin and can extend over a broader frequency range than $S^{E}(\Omega)$. In particular, if the field is in a coherent state (including the vacuum) one can show by using (\ref{commut}) that:
\begin{eqnarray}\label{vide}
{S^{IN}(\Omega)}_{\mathit{Coh}}=\frac{\hbar \omega_{L}}{2\epsilon_{0} Ac}\left( \begin{matrix}
1 & 0\\
0 & 0
\end{matrix} \right).
\end{eqnarray}

\subsection{Observables}
Starting from the two field operator correlations, it is possible to compute several observables connected to the spectral correlation matrix.
\subsubsection{Optical spectrum}
First, the optical spectrum of the field, which can be measured using a spectrometer, is given by:
\begin{eqnarray}
S_{Opt}(\omega_L+\Omega)\propto\int e^{i\Omega\tau} \langle a^{\dagger}(\tau) a(0)\rangle d\tau \propto\mathcal{S}{(\vert\Omega\vert)}_{22}.
\end{eqnarray}
This optical spectrum includes an elastic component at the driving frequency of the incident laser, as one would also expect in linear optics of driven harmonic oscillators. As we will see below, quantum features appear in the optical spectrum, with the Mollow triplet considered as a hallmark of quantum fluctuations \cite{MOLLOW69}, emerging for strong driving field. We will also see that the multilevel transition leaves an imprint in the optical spectrum even at low incident intensities where the Mollow triplet contribution is very small in comparison to the elastic scattering.

\subsubsection{Field quadratures}
From the spectral correlation matrix, one can also compute the noise for the field quadratures. A generalized field quadrature (for a given polarization component) is defined as:
\begin{eqnarray}
X_\theta(t)=\xi \left[ a(t)e^{-i\theta}+a^{\dagger}(t)e^{i\theta} \right].
\end{eqnarray}
Field quadratures can be measured using a homodyne detection system in which the angle $\theta$ corresponds to the phase of the local oscillator~\cite{Bimbard_Homodyne_14}. Moreover, when an intense field is directly incident on a photodetector, the photocurrent is given by:
\begin{eqnarray}\label{PD}
I(t)\propto a(t)^{\dagger}a(t)\simeq \vert \langle a\rangle \vert^{2} +\langle a\rangle^{*}\delta a(t)+\langle a\rangle\delta a(t)^{\dagger}.
\end{eqnarray}
If the variations of the mean value of the field can be neglected (or filtered out) in the frequency range of interest, then the signal fluctuations are dominated by the last two terms in (\ref{PD}), which correspond to a quadrature operator $X_{\theta}$ whose angle $\theta$ is given by the phase of the field mean value.\\

The quadrature noise spectrum of a stationary field reads:
\begin{eqnarray}
S_{X_\theta}(\Omega)&=&\xi^{2}\int e^{i\Omega\tau} \langle \delta X_{\theta}(\tau)\delta X_{\theta}(0)\rangle d\tau\nonumber\\
&=& \xi^{2}\left[\mathcal{S}_{11}+\mathcal{S}_{12}e^{-2i\theta}+\mathcal{S}_{21}e^{2i\theta}+\mathcal{S}_{22}\right].
\end{eqnarray}\\

Both $S_{Opt}$ and $S_{X_\theta}$ include contributions from $S^{E}$ and $S^{IN}$. In the experimental observation of the optical spectrum, the elastic contribution to the spectrum is often dominant, specially at low light intensity and it is challenging to extract the inelastic contribution. On the other hand, when quadrature noise fluctuations are measured, the spectral range corresponding to the contribution from $S^{E}(\Omega)$ is commonly avoided since it is dominated by technical noise coming from the light source itself. In this article we are mainly concerned with the inelastic contribution to the fluctuation spectra.

\subsection{Outline of the calculation}

The numerical calculation of the spectral correlation matrix $S^{IN}(\Omega)$ for light that has passed through an atomic sample was previously presented in ~\cite{LEZAMA08}. We here recall the essential lines and hypothesis of this calculation, the reader is referred to ~\cite{LEZAMA08} for all the details. The atomic system is considered as an homogeneous sample of atoms, with ground state total angular momentum $F_g$ and excited state angular momentum $F_e$. The energy of the transition is $\hbar \omega_0$, and the decay rate is $\Gamma$. The levels are Zeeman degenerate, with $2F_g+1$ (resp. $2F_e+1$) Zeeman sublevels in the ground (resp. excited) state. Individual states are labelled $\vert F,M_F\rangle$, where $M_F$ is the magnetic quantum number. The atomic operators are of the form $\sigma_{\alpha,\beta}\equiv\vert \alpha\rangle\langle \beta\vert$, where $\alpha$ and $\beta$ designate an ($F, M_F$) pair.

\subsubsection{Atomic evolution}

Let $\sigma=\left\lbrace \sigma_{\alpha,\beta}(z) \right\rbrace $ represent the ensemble of the atomic operators as a function of the position $z$~\cite{DANTAN05}. In the presence of the optical field, the atomic operators evolve according to Heisenberg-Langevin equations. These equations which are first order linear differential equations for the operators $\sigma_{\alpha,\beta}(z)$ can be cast into the form:
\begin{eqnarray}\label{HL}
\frac{d\sigma}{dt}=\mathsf{H_0}(\sigma)+\mathsf{V}(\sigma)+\mathsf{R}(\sigma)+f,
\end{eqnarray}
where $\mathsf{H_0}$, $\mathsf{V}$ and $\mathsf{R}$ are linear operators acting upon $\sigma$. $\mathsf{H_0}$ represents the free Hamiltonian evolution. $\mathsf{V}$ is the dipolar light-atom interaction coupling (in the rotating wave approximation), which depends linearly on the operators $a_1$, $a_2$ and their Hermitian conjugates. $\mathsf{R}$ describes the atomic relaxation, including all possible spontaneous decay channels between Zeeman sublevels. The general expressions of $\mathsf{H_0}$, $\mathsf{V}$ and $\mathsf{R}$ are given in~\cite{LEZAMA08} . $f=\left\lbrace f_{\alpha,\beta} \right\rbrace$ represent the ensemble of Langevin forces which have zero mean value and satisfy~\cite{DANTAN05}:
\begin{equation}
\left\langle f_{\alpha \beta} (z,t) f_{\gamma \delta}^{\dagger } (z^{\prime
},t^{\prime }) \right\rangle =\frac LN2D_{\alpha \beta,\gamma \delta}\delta
 (z^{\prime }-z^{\prime \prime }) \delta (t-t^{\prime })
\end{equation}
where $N$ is the total number of atoms, $D_{\alpha \beta,\gamma \delta}$ is the corresponding diffusion coefficient. 
 
The diffusion coefficients that depend on the specific transition and the incident light intensity and detuning, are numerically calculated from Eq. (\ref{HL}) using the generalized Einstein theorem \cite{SARGENTBOOK74,Cohen1992} following the procedure described in ~\cite{LEZAMA08}.
The mean value of Eq.~(\ref{HL}) corresponds to the usual optical Bloch equations, which are used to calculate the atomic density matrix $\rho$ in the presence of the field~\cite{Comment2}.

\subsubsection{Field evolution}
The field evolution is governed by the Maxwell-Heisenberg equations (in the slowly varying envelope approximation):

\begin{eqnarray}\label{Maxwell}
\left ( \frac \partial {\partial t}+c\frac \partial {\partial z} \right ) a_\lambda &=&iN\eta p_\lambda(z),
\end{eqnarray}
with $\lambda=1,2$. Here $p_\lambda(z)$ is the rising atomic operator projected along the polarization $\hat{e}_\lambda$ (it is a linear combination of the atomic operators $\sigma_{\alpha,\beta}$ involving Clebsch-Gordan coefficients, see Eqs. 6 in ~\cite{LEZAMA08}) and $\eta=\xi\mu/\hbar $ is the atom-photon coupling constant ($\mu = \left\langle g\Vert \vec{D}\Vert e\right\rangle$ is the \emph{reduced} matrix element of the atomic dipole operator).

\subsubsection{Calculation procedure}
We consider simultaneously the evolution of two orthogonal polarization modes of the field therefore including a possible coupling of the two components by the atomic system. The calculation proceeds through the following steps:
\begin{enumerate}[a)]
\item Linearisation of the atomic and field observables;
\item Fourier transformation of Eqs. (\ref{HL}) and (\ref{Maxwell}).
\item Evaluation from Eq.~\ref{HL} of the atomic operators fluctuations at a given position $z$ as a function of the Langevin forces. Two important simplifications are introduced at this stage. First, the mean values of the field are assumed to be independent of $z$ and consequently the mean value of the atomic operators are also independent of $z$. Moreover, the effect of the field fluctuations on the atomic fluctuations is neglected.
\item Substitution of the atomic fluctuations into Eqs.~(\ref{Maxwell}), which become propagations equations for the field fluctuations driven by the atomic fluctuations.
\item Formal integration of the field fluctuation propagation equations.
\item Calculation of the expectation value of the two-frequencies field fluctuations products as a function of the Langevin forces diffusion coefficients.  
\item Derivation the expression $S^{IN}(\Omega,L)$ in terms of the incident spectral correlation matrix $S^{IN}(\Omega,z=0)$ ($L$ is the length of the atomic medium).
\end{enumerate}

\subsubsection{Parameters}
The parameters used in the calculation are: the detuning between the laser and atomic resonance frequency $\Delta\equiv\omega_L-\omega_0$, the Rabi frequency $\Omega_1\equiv\mu E_1/\hbar$ of the incident field component with polarization $\hat{e}_1$ ($E_1$ is the field amplitude) and the \emph{reduced} on-resonance optical density of the atomic sample $$b_0=\frac{2\rho \omega_L \mu^2Z}{\hbar\epsilon_0 c\Gamma} =\dfrac{3\lambda^{2}}{2\pi}\rho L,$$ where $\rho$ is the atomic density and $\lambda$ the light wavelength in vacuum. Notice that the reduced on-resonance optical density differs from the actual optical density of the sample by a numerical factor depending on the light polarization and detuning.

\subsubsection{State of the incident field}
The initial value $S^{IN}(\Omega,z=0)$ for the intense field polarization component $\hat{e}_1$ is taken in the form:
\begin{eqnarray}\label{exceso}
S^{IN}=\frac{\hbar \omega_{L}}{2\epsilon_{0} Ac}\left[ \left( \begin{matrix}
1 & 0\\
0 & 0
\end{matrix} \right)+\frac{1}{4}\left( \begin{matrix}
\varepsilon_A+\varepsilon_P & \varepsilon_A-\varepsilon_P\\
\varepsilon_A-\varepsilon_P & \varepsilon_A+\varepsilon_P
\end{matrix} \right)\right]
\end{eqnarray}

The second term in (\ref{exceso}) describes \emph{white, uncorrelated} excess noise in the incident field quadratures above the level of vacuum fluctuations. $\varepsilon_A$ and $\varepsilon_P$ represent the fractional excess relative to the vacuum noise level of the variances of the amplitude ($\theta=0$) and phase ($\theta=\pi/2$) field quadrature fluctuations respectively. We always use $\varepsilon_A=\varepsilon_P =0$ for the polarization component perpendicular to the incident field.\\
\section{Results} 

To address the problem considered in this paper, we have computed the field fluctuations for two different incident beam polarization modes, applied to an ensemble of atoms having a near resonant $F_g=1\rightarrow F_e=2$ transition (see \figurename{~\ref{levels}}). \figurename{~\ref{levels}}(a) corresponds to an incident field with circular polarization. Due to the Zeeman optical pumping, the steady state population of the atomic system is restricted to the states $\vert F_g=1,m=1\rangle$ and $\vert F_e=2,m=2 \rangle$, thus effectively realizing a two-level system (TLS). No light is scattered by the atomic system into the orthogonal (circular) field polarization. \figurename{~\ref{levels}}(b) corresponds to linear polarization, where all three ground level Zeeman states are populated at steady state and connected to the excited states through the $\Delta m=0$ selection rule. Therefore, this case corresponds to a multi-level system (MLS), for which the atoms are coupled to both orthogonal polarization components of the field.

\subsection{Optical spectrum}
We initially compute the \textit{inelastic} contribution to the optical spectrum for the TLS and for the MLS. This problem has previously been examined by Bo Gao~\cite{Gao1994}. We here consider an atomic sample with optical density $b_0=0.1$ and different choices of the incident field Rabi frequency. We thus avoid the regime of multiple scattering with an important attenuation of the incident laser field. We initially assume that the incident field is in a coherent state ($\varepsilon_A=\varepsilon_P=0$).\\

\begin{figure}
\includegraphics[width=8.5cm]{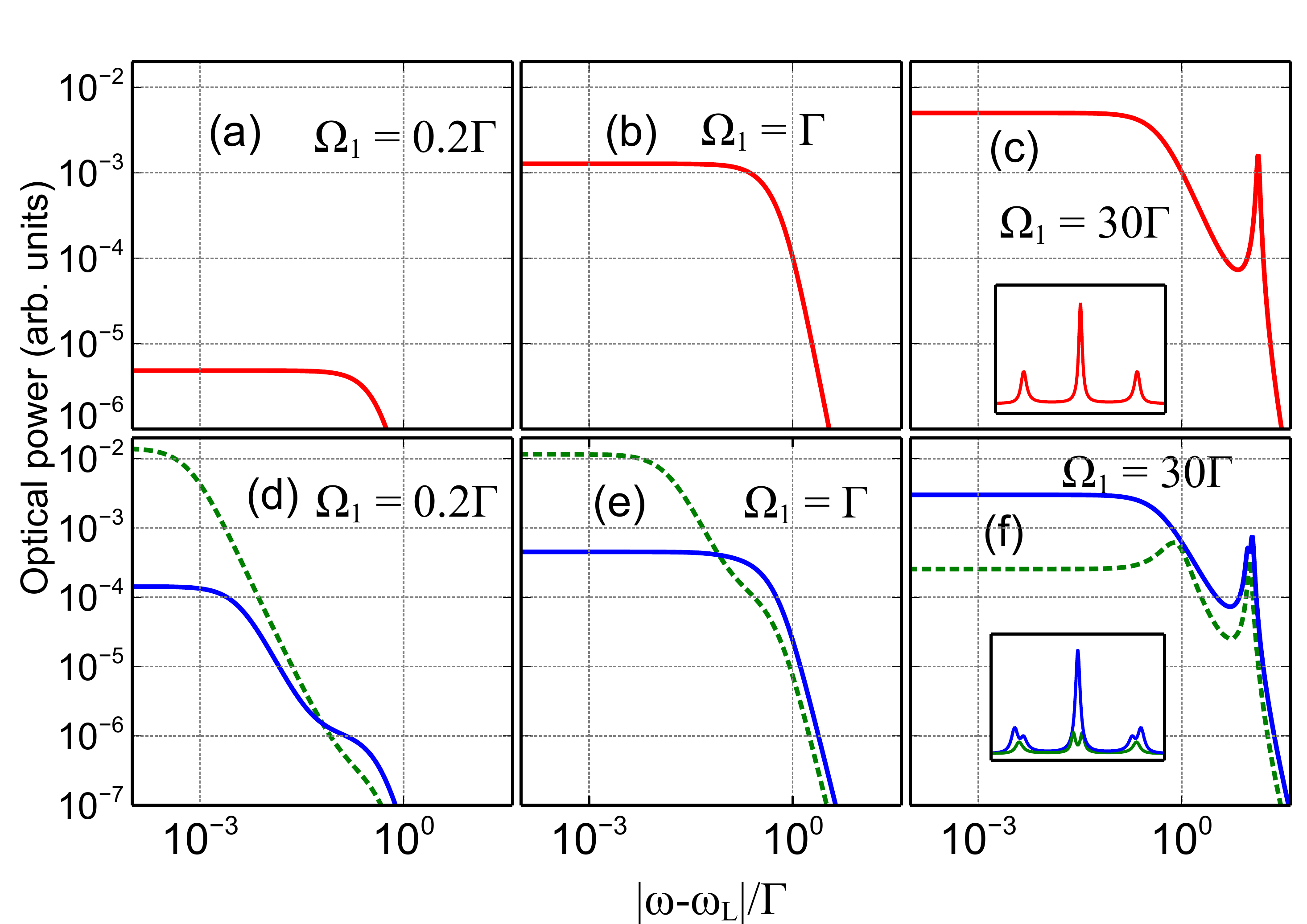}
\caption{\label{optical}(Color online) Inelastic part of the optical spectrum of resonant light after traversing the atomic sample, as a function of the frequency offset $\vert\omega-\omega_L\vert$ for different values of the incident field Rabi frequency $\Omega_1$. Upper row: TLS. Lower row: MLS. Solid lines: incident field polarization. Dashed: orthogonal polarization. Insets: same spectra on a linear scale (symmetrized around $\omega-\omega_L=0$). [$\Delta=0$, $b_0=0.1$, $\varepsilon_A=\varepsilon_P=0$].}
\end{figure}

The computed inelastic part of the optical spectrum at angular frequency $\omega$ is presented in \figurename{~\ref{optical}}, as a function of $\vert \omega - \omega_L \vert$ (the spectra are symmetric around $\omega_L$), for three choices of incident field Rabi frequency and exact resonance of the light frequency with the atomic transition. The plots are presented in log-log scales in order to stress the different behaviors depending on the magnitude of $\vert \omega - \omega_L \vert$.

The upper row in \figurename{~\ref{optical}} corresponds to the TLS. It reproduces the Mollow triplet spectrum~\cite{MOLLOW69}. For low Rabi frequencies, the spectrum consist of a single peak with a width of the order of $\Gamma$, the natural linewidth of the transition. It is worth reminding at this point that in this regime the elastic contribution to the spectrum (not presented in \figurename{~\ref{optical}}) is dominant \cite{Cohen1992}. For Rabi frequencies $\Omega_1$ larger than $\Gamma$, the spectrum evolves into the characteristic triplet structure (only the positive half is shown) where the side-bands are separated from the center by $\zeta\Omega_1$ ($\zeta$ is the coupling Clebsh-Gordan coefficient for the transition). The widths of all peaks are on the order of $\Gamma$.\\

The second row in \figurename{~\ref{optical}} corresponds to the MLS case. As expected, the inelastic part of the optical spectrum here contains contributions from the two optical polarizations components (albeit only the incident field polarization contributes to the elastic part of the optical spectrum). The most significant difference in the spectra compared to the TLS case appears for low Rabi frequencies, where, for both polarizations, the spectra present peaks at the origin, with widths on the order of $\Omega_1^2/\Gamma$. The origin of these peaks is the spontaneous Raman scattering of light into the $\Delta m=\pm 1$ transitions~\cite{Gao1994}, which is not present in the TLS. Note that the height of the peak is significantly larger for the orthogonal polarization than for the incident polarization.

Other differences between the MLS and TLS optical spectra arise for large Rabi frequencies. The incident field polarization spectrum evolves into a quintuplet, while the orthogonal polarization spectrum becomes a quadruplet. Both structures can be easily understood as a consequence of the different light-shifts of the various $\Delta m=0$ transitions due to differences in the corresponding coupling coefficients \cite{Lezama1999}.

\subsection{Quadrature noise spectrum}

We now examine the spectrum of the noise in the field quadrature. As before, we only consider the inelastic part of the spectrum and we limit the study to the amplitude quadrature,
since it is in principle readily accessible by direct photodetection of the total light intensity transmitted through a cloud of cold atoms at low optical thickness. Indeed, in this limit, the unscattered incident field acts as a local oscillator, which perfect mode matching in the forward direction.

\begin{figure}
\includegraphics[width=8.5cm]{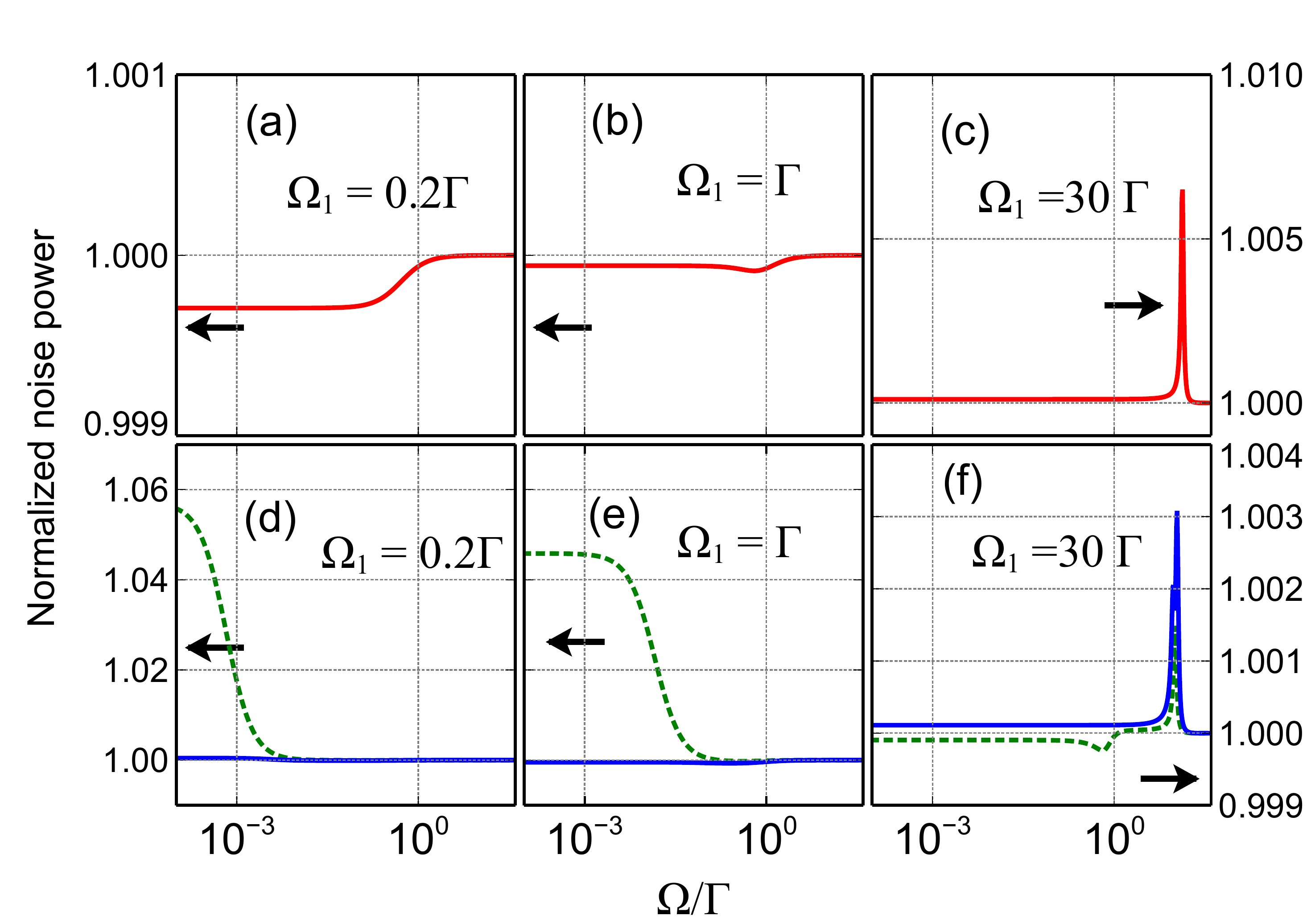}
\caption{\label{noise}(Color online) Inelastic part of the noise spectrum in the amplitude quadrature of resonant light after traversing the atomic sample, as a function of the noise frequency $\Omega$ for different values of the incident field Rabi frequency $\Omega_1$. The noise power is normalized by the shot noise level. Upper row: TLS. Lower row: MLS. Solid: incident field polarization. Dashed: orthogonal polarization. [$\Delta=0$, $b_0=0.1$, $\varepsilon_A=\varepsilon_P=0$]. The arrows indicate the vertical axes, corresponding to the respective spectra.}
\end{figure}

The computed noise spectra are presented in \figurename{~\ref{noise}}, for the same conditions as those considered in \figurename{~\ref{optical}}, as a function of the noise frequency $\Omega$. Note that the spectra are now presented in a semi-log scale. The noise power is normalized to the power in the vacuum mode (referred to as shot noise level). For the TLS case, at low noise frequencies, the quadrature fluctuations are squeezed over a frequency range on the order of $\Gamma$~\cite{COLLETT84,HEIDMANN85,HO87}. The squeezing initially increases with the Rabi frequency for $\Omega_1\lesssim \Gamma$ and then decreases for $\Omega_1 > \Gamma$. At large Rabi frequencies an excess noise peak appears around $\Omega=\zeta\Omega_1$, due to spontaneous emission on the Mollow triplet sidebands.\\

Significant qualitative differences appear in the noise spectrum for the MLS case. At low noise frequency, instead of squeezing, we observe an excess noise peak for both field polarizations. For the incident peak polarization, the amplitude of the peak is too small to be appreciated on the scale of \figurename{~\ref{noise}}(d,e). As for the optical spectrum, the low frequency peak has a width of the order of $\Omega_1^2/\Gamma$. At large Rabi frequencies, excess noise peaks appear for both polarizations around $\Omega\simeq\Omega_1$, due to the spontaneous emission on the resonance fluorescence sidebands. Also, the quadrature noise on the orthogonal polarization is squeezed for noise frequencies below $\Omega_1$.

\begin{figure}
\includegraphics[width=8.5cm]{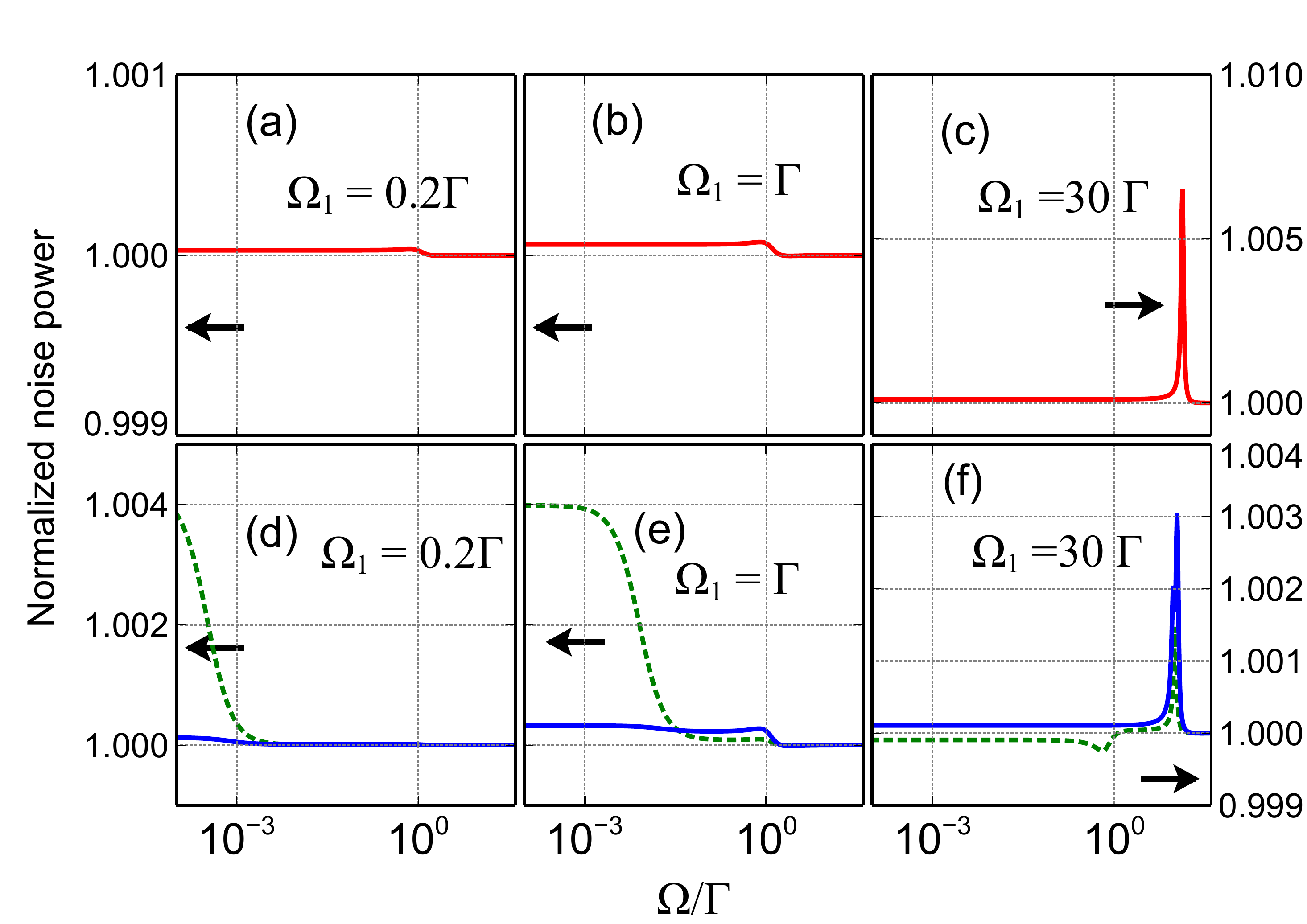}
\caption{\label{noisedet}(Color online) Same as in \figurename{~\ref{noise}}, but with a detuned incident laser field ($\Delta=\Gamma$).}
\end{figure}

The induced quadrature noise when the laser is detuned from resonance by $\omega-\omega_L=\Gamma$ is presented in \figurename{~\ref{noisedet}}. Note that the squeezing is suppressed for the TLS case. For the MLS case, a noticeable increase in excess noise occurs for the incident field polarization, in addition to the appearance of small features around $\Omega=\Delta$. The noise peak around $\Omega=0$ is narrower than in the case of the resonant excitation for both polarization components.

\subsection{Effect of the laser noise}
Usually, actual laser beams have noise specifications different than those of a coherent state. In particular, diode lasers are known to operate at (or even below) the shot noise level regarding the intensity, while possessing relatively large frequency (or phase) noise. If such an excess noise is not too large, it can be approximately described as phase quadrature excess noise.

To take into account the commonly encountered experimental condition of excess laser phase noise, we have computed the value of the amplitude quadrature noise of the transmitted field, assuming an incident field excess quadrature noise described by $\varepsilon_P >0$ (keeping $\varepsilon_A=0$). We consider the case of a detuned laser field ($\Delta=\Gamma$) since in this case the mean atomic dipole has a non-zero in-phase component that will transform phase quadrature noise into amplitude quadrature noise.\\

\begin{figure}
\includegraphics[width=8.5cm]{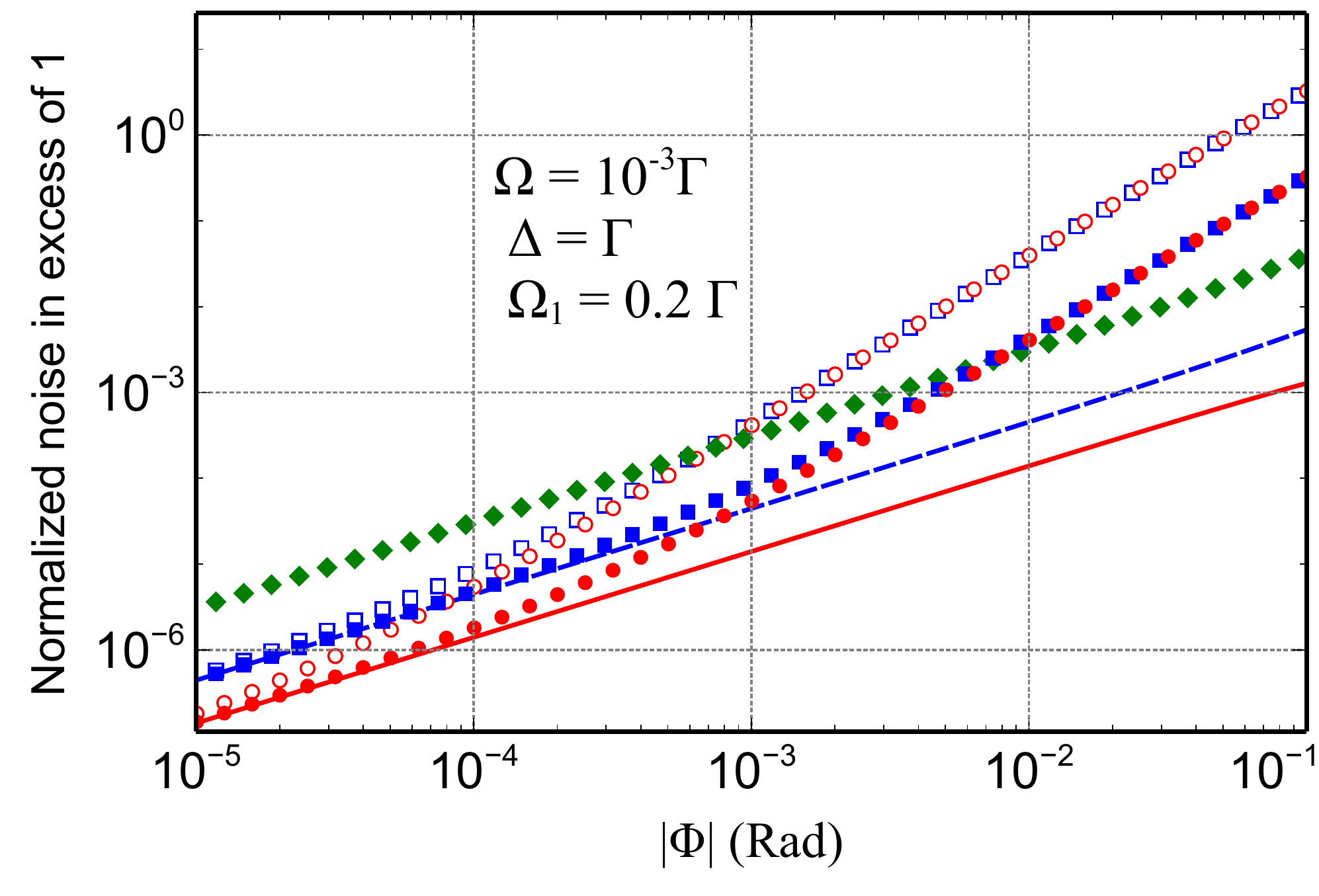}
\caption{\label{noisedetalpha}(Color online) Quadrature noise power as a function of the dephasing angle $\Phi$ introduced by the atomic sample. Lines: TLS (solid) and MLS incident field polarization (dashed) with no incident field excess noise ($\varepsilon_P=0$). Circles: TLS. Squares: MLS for the incident field polarization. Filled symbols: $\varepsilon_P=10$, hollow symbols: $\varepsilon_P=100$. Diamonds: MLS orthogonal polarization for all values of $\varepsilon_P$.}
\end{figure}

Figure~\ref{noisedetalpha} shows the calculated noise power at a noise frequency $\Omega=10^{-3}\Gamma$, for an incident Rabi frequency $\Omega_1=0.2\Gamma$, as a function of the absolute value of the field dephasing $\Phi\equiv b_0 Real(Tr[\rho p_1(z)]) L/4\Omega_1$ introduced by the propagation through the atomic sample for three values of $\varepsilon_P=0,10,100$. Since the incident field parameters are given, the value of $\Phi$ essentially reflects its linear dependence on the reduced on-resonance optical density $b_0$. However, the factor $Real(Tr[\rho p_1(z)])$ incorporates the dependence of the atomic response on the incident field polarization, intensity and detuning.

The lines in \figurename{~\ref{noisedetalpha}} correspond to $\varepsilon_P=0$ for both the TLS and MLS cases. The dependence of the noise on $b_0$, and therefore also on $\vert\Phi\vert$, is linear. The symbols in \figurename{~\ref{noisedetalpha}} correspond to situations where the excess noise is present. The amplitude quadrature noise for the incident field polarization increases for both the TLS and MLS in proportion to $\varepsilon_P$. Notice that for large $\vert\Phi\vert$, the amplitude quadrature noise now increases quadratically with $b_0$. There is no influence of $\varepsilon_P$ on the amplitude quadrature noise in the orthogonal polarization in the MLS.

The main result arising from \figurename{~\ref{noisedetalpha}} is the classical behavior of the amplitude quadrature noise in the presence of excess noise for sufficiently large optical density. In that regime, the quadrature noise is essentially the same for the TLS and MLS since it is dominated by classical conversion of the phase excess noise into amplitude noise via the mean value of the atomic sample polarization. By considering different values of the optical detuning and the Rabi frequency, we have checked that the the classical phase to amplitude noise conversion mechanism mainly depends on these parameters through the phase angle $\Phi$. Only in the limit of very small $\vert\Phi\vert$ the qualitative differences between the TLS and the MLS noise properties are observable.

\subsection{Conclusion}
In this paper, we report on the study of optical and quadrature noise spectra, comparing two-level to multilevel transitions. We have shown that the presence of spontaneous Raman transition produces additional components both in the optical spectrum, as previously studied by Bo Gao \cite{Gao1993, Gao1994}, as well as in the low frequency noise spectrum of the light transmitted through a sample of cold atoms. These results confirm the fundamental role of the Zeeman degeneracy on the fluctuations in light-atom interactions. We have also considered the realistic case of additional frequency noise of the incident laser which has to be taken into account when comparing the results presented in this paper to experiments using clouds of laser cooled atoms. The approach used in this paper can readily be extended to study the additional noise in room temperature atomic vapors \cite{LEZAMA08}, which requires to include Doppler broadening as well as multiple transitions, relevant in Doppler broadened samples. Going beyond this approach for exploring the regime of multiple scattering and the noise or correlation functions in the field scattered out of the incident laser mode, would allow to confront the predictions in~\cite{Akkermans2007,Gremaud2008,Akkermans2008} to a microscopic \textit{ab initio} model as used in this work.\\

\acknowledgments
This work was supported by the ECOS grant U14E01 and by the F\'ed\'eration Doeblin (CNRS FR 2800) and CSIC (Uruguay).

%

\begin{thebibliography}{40}
\expandafter\ifx\csname natexlab\endcsname\relax\def\natexlab#1{#1}\fi
\expandafter\ifx\csname bibnamefont\endcsname\relax
  \def\bibnamefont#1{#1}\fi
\expandafter\ifx\csname bibfnamefont\endcsname\relax
  \def\bibfnamefont#1{#1}\fi
\expandafter\ifx\csname citenamefont\endcsname\relax
  \def\citenamefont#1{#1}\fi
\expandafter\ifx\csname url\endcsname\relax
  \def\url#1{\texttt{#1}}\fi
\expandafter\ifx\csname urlprefix\endcsname\relax\def\urlprefix{URL }\fi
\providecommand{\bibinfo}[2]{#2}
\providecommand{\eprint}[2][]{\url{#2}}

\bibitem[{\citenamefont{Chu}(1998)}]{Chu1998}
\bibinfo{author}{\bibfnamefont{S.}~\bibnamefont{Chu}}, \bibinfo{journal}{Rev.
  Mod. Phys.} \textbf{\bibinfo{volume}{70}}, \bibinfo{pages}{685}
  (\bibinfo{year}{1998}).

\bibitem[{\citenamefont{Cohen-Tannoudji}(1998)}]{CohenTannoudji1998}
\bibinfo{author}{\bibfnamefont{C.}~\bibnamefont{Cohen-Tannoudji}},
  \bibinfo{journal}{Rev. Mod. Phys.} \textbf{\bibinfo{volume}{70}},
  \bibinfo{pages}{707} (\bibinfo{year}{1998}).

\bibitem[{\citenamefont{Philipps}(1998)}]{Philipps1998}
\bibinfo{author}{\bibfnamefont{W.~D.} \bibnamefont{Philipps}},
  \bibinfo{journal}{Rev. Mod. Phys.} \textbf{\bibinfo{volume}{70}},
  \bibinfo{pages}{721} (\bibinfo{year}{1998}).

\bibitem[{\citenamefont{Lvovsky}(2015)}]{Squeezing}
\bibinfo{author}{\bibfnamefont{A.~I.} \bibnamefont{Lvovsky}},
  \emph{\bibinfo{title}{Squeezed Light}} (\bibinfo{publisher}{John Wiley \&
  Sons, Inc.}, \bibinfo{year}{2015}), pp. \bibinfo{pages}{121--163}, ISBN
  \bibinfo{isbn}{9781119009719},
  \urlprefix\url{http://dx.doi.org/10.1002/9781119009719.ch5}.

\bibitem[{\citenamefont{Lambrecht et~al.}(1996)\citenamefont{Lambrecht,
  Coudreau, Steinberg, and Giacobino}}]{LAMBRECHT96}
\bibinfo{author}{\bibfnamefont{A.}~\bibnamefont{Lambrecht}},
  \bibinfo{author}{\bibfnamefont{T.}~\bibnamefont{Coudreau}},
  \bibinfo{author}{\bibfnamefont{A.~M.} \bibnamefont{Steinberg}},
  \bibnamefont{and}
  \bibinfo{author}{\bibfnamefont{E.}~\bibnamefont{Giacobino}},
  \bibinfo{journal}{EPL (Europhysics Letters)} \textbf{\bibinfo{volume}{36}},
  \bibinfo{pages}{93} (\bibinfo{year}{1996}),
  \urlprefix\url{http://stacks.iop.org/0295-5075/36/i=2/a=093}.

\bibitem[{\citenamefont{Ries et~al.}(2003)\citenamefont{Ries, Brezger, and
  Lvovsky}}]{Ries03}
\bibinfo{author}{\bibfnamefont{J.}~\bibnamefont{Ries}},
  \bibinfo{author}{\bibfnamefont{B.}~\bibnamefont{Brezger}}, \bibnamefont{and}
  \bibinfo{author}{\bibfnamefont{A.~I.} \bibnamefont{Lvovsky}},
  \bibinfo{journal}{Phys. Rev. A} \textbf{\bibinfo{volume}{68}},
  \bibinfo{pages}{025801} (\bibinfo{year}{2003}),
  \urlprefix\url{http://link.aps.org/doi/10.1103/PhysRevA.68.025801}.

\bibitem[{\citenamefont{Labeyrie et~al.}(2003)\citenamefont{Labeyrie, Delande,
  Mueller, Miniatura, and Kaiser}}]{Labeyrie2003}
\bibinfo{author}{\bibfnamefont{G.}~\bibnamefont{Labeyrie}},
  \bibinfo{author}{\bibfnamefont{D.}~\bibnamefont{Delande}},
  \bibinfo{author}{\bibfnamefont{C.~A.} \bibnamefont{Mueller}},
  \bibinfo{author}{\bibfnamefont{C.}~\bibnamefont{Miniatura}},
  \bibnamefont{and} \bibinfo{author}{\bibfnamefont{R.}~\bibnamefont{Kaiser}},
  \bibinfo{journal}{Europhys. Lett.} \textbf{\bibinfo{volume}{61}},
  \bibinfo{pages}{327} (\bibinfo{year}{2003}).

\bibitem[{\citenamefont{Assaf and Akkermans}(2007)}]{Akkermans2007}
\bibinfo{author}{\bibfnamefont{O.}~\bibnamefont{Assaf}} \bibnamefont{and}
  \bibinfo{author}{\bibfnamefont{E.}~\bibnamefont{Akkermans}},
  \bibinfo{journal}{Phys. Rev. Lett.} \textbf{\bibinfo{volume}{98}},
  \bibinfo{pages}{083601} (\bibinfo{year}{2007}).

\bibitem[{\citenamefont{Gr\'emaud et~al.}(2008)\citenamefont{Gr\'emaud,
  Delande, M\"uller, and Miniatura}}]{Gremaud2008}
\bibinfo{author}{\bibfnamefont{B.}~\bibnamefont{Gr\'emaud}},
  \bibinfo{author}{\bibfnamefont{D.}~\bibnamefont{Delande}},
  \bibinfo{author}{\bibfnamefont{C.}~\bibnamefont{M\"uller}}, \bibnamefont{and}
  \bibinfo{author}{\bibfnamefont{C.}~\bibnamefont{Miniatura}},
  \bibinfo{journal}{Phys. Rev. Lett.} \textbf{\bibinfo{volume}{100}},
  \bibinfo{pages}{199301} (\bibinfo{year}{2008}).

\bibitem[{\citenamefont{Assaf and Akkermans}(2008)}]{Akkermans2008}
\bibinfo{author}{\bibfnamefont{O.}~\bibnamefont{Assaf}} \bibnamefont{and}
  \bibinfo{author}{\bibfnamefont{E.}~\bibnamefont{Akkermans}},
  \bibinfo{journal}{Phys. Rev. Lett.} \textbf{\bibinfo{volume}{100}},
  \bibinfo{pages}{199302} (\bibinfo{year}{2008}).

\bibitem[{\citenamefont{M\"uller et~al.}(2015)\citenamefont{M\"uller,
  Gr\'emaud, and Miniatura}}]{Miniatura2015}
\bibinfo{author}{\bibfnamefont{C.~A.} \bibnamefont{M\"uller}},
  \bibinfo{author}{\bibfnamefont{B.}~\bibnamefont{Gr\'emaud}},
  \bibnamefont{and}
  \bibinfo{author}{\bibfnamefont{C.}~\bibnamefont{Miniatura}},
  \bibinfo{journal}{Phys. Rev. A} \textbf{\bibinfo{volume}{92}},
  \bibinfo{pages}{013819} (\bibinfo{year}{2015}).

\bibitem[{\citenamefont{Polder and Schuurmans}(1976)}]{Polder1976}
\bibinfo{author}{\bibfnamefont{D.}~\bibnamefont{Polder}} \bibnamefont{and}
  \bibinfo{author}{\bibfnamefont{M.}~\bibnamefont{Schuurmans}},
  \bibinfo{journal}{Phys. Rev. A} \textbf{\bibinfo{volume}{14}},
  \bibinfo{pages}{1468} (\bibinfo{year}{1976}).

\bibitem[{\citenamefont{Cooper et~al.}(1980)\citenamefont{Cooper, Ballagh, and
  Burnett}}]{Cooper1980}
\bibinfo{author}{\bibfnamefont{J.}~\bibnamefont{Cooper}},
  \bibinfo{author}{\bibfnamefont{R.}~\bibnamefont{Ballagh}}, \bibnamefont{and}
  \bibinfo{author}{\bibfnamefont{K.}~\bibnamefont{Burnett}},
  \bibinfo{journal}{Phys. Rev. A} \textbf{\bibinfo{volume}{22}},
  \bibinfo{pages}{535} (\bibinfo{year}{1980}).

\bibitem[{\citenamefont{Javanainen}(1992)}]{Javanainen1992}
\bibinfo{author}{\bibfnamefont{J.}~\bibnamefont{Javanainen}},
  \bibinfo{journal}{EPL} \textbf{\bibinfo{volume}{20}}, \bibinfo{pages}{395}
  (\bibinfo{year}{1992}).

\bibitem[{\citenamefont{Gao}(1993)}]{Gao1993}
\bibinfo{author}{\bibfnamefont{B.}~\bibnamefont{Gao}}, \bibinfo{journal}{Phys.
  Rev. A} \textbf{\bibinfo{volume}{48}}, \bibinfo{pages}{2443}
  (\bibinfo{year}{1993}).

\bibitem[{\citenamefont{Gao}(1994)}]{Gao1994}
\bibinfo{author}{\bibfnamefont{B.}~\bibnamefont{Gao}}, \bibinfo{journal}{Phys.
  Rev. A} \textbf{\bibinfo{volume}{50}}, \bibinfo{pages}{4139}
  (\bibinfo{year}{1994}).

\bibitem[{\citenamefont{Galbraith et~al.}(1982)\citenamefont{Galbraith, Dubs,
  and Steinfeld}}]{Galbraith1982}
\bibinfo{author}{\bibfnamefont{H.}~\bibnamefont{Galbraith}},
  \bibinfo{author}{\bibfnamefont{M.}~\bibnamefont{Dubs}}, \bibnamefont{and}
  \bibinfo{author}{\bibfnamefont{J.}~\bibnamefont{Steinfeld}},
  \bibinfo{journal}{Phys. Rev. A} \textbf{\bibinfo{volume}{26}},
  \bibinfo{pages}{1528} (\bibinfo{year}{1982}).

\bibitem[{\citenamefont{Berman}(2008)}]{Berman2008}
\bibinfo{author}{\bibfnamefont{P.}~\bibnamefont{Berman}},
  \bibinfo{journal}{Contemporary Physics} \textbf{\bibinfo{volume}{49}},
  \bibinfo{pages}{313} (\bibinfo{year}{2008}).

\bibitem[{\citenamefont{van Bergen et~al.}(1988)\citenamefont{van Bergen, van
  Halewijn, Hollander, and Alkemade}}]{Bergen1988}
\bibinfo{author}{\bibfnamefont{A.~R.~D.} \bibnamefont{van Bergen}},
  \bibinfo{author}{\bibfnamefont{H.~J.} \bibnamefont{van Halewijn}},
  \bibinfo{author}{\bibfnamefont{T.}~\bibnamefont{Hollander}},
  \bibnamefont{and} \bibinfo{author}{\bibfnamefont{C.~T.~J.}
  \bibnamefont{Alkemade}}, \bibinfo{journal}{J. Phys. B}
  \textbf{\bibinfo{volume}{21}}, \bibinfo{pages}{647} (\bibinfo{year}{1988}).

\bibitem[{\citenamefont{Lezama et~al.}(1999)\citenamefont{Lezama, Barreiro,
  Lipsich, and Akulshin}}]{Lezama1999}
\bibinfo{author}{\bibfnamefont{A.}~\bibnamefont{Lezama}},
  \bibinfo{author}{\bibfnamefont{S.}~\bibnamefont{Barreiro}},
  \bibinfo{author}{\bibfnamefont{A.}~\bibnamefont{Lipsich}}, \bibnamefont{and}
  \bibinfo{author}{\bibfnamefont{A.}~\bibnamefont{Akulshin}},
  \bibinfo{journal}{Physical Review A} \textbf{\bibinfo{volume}{61}},
  \bibinfo{pages}{013801} (\bibinfo{year}{1999}).

\bibitem[{\citenamefont{Walker et~al.}(1996)\citenamefont{Walker, Bali,
  Hoffmann, and Sima}}]{Walker1996}
\bibinfo{author}{\bibfnamefont{T.~G.} \bibnamefont{Walker}},
  \bibinfo{author}{\bibfnamefont{S.}~\bibnamefont{Bali}},
  \bibinfo{author}{\bibfnamefont{D.}~\bibnamefont{Hoffmann}}, \bibnamefont{and}
  \bibinfo{author}{\bibfnamefont{J.}~\bibnamefont{Sima}},
  \bibinfo{journal}{Phys. Rev. A} \textbf{\bibinfo{volume}{53}},
  \bibinfo{pages}{2} (\bibinfo{year}{1996}).

\bibitem[{\citenamefont{Lezama et~al.}(2008)\citenamefont{Lezama, Valente,
  Failache, Martinelli, and Nussenzveig}}]{LEZAMA08}
\bibinfo{author}{\bibfnamefont{A.}~\bibnamefont{Lezama}},
  \bibinfo{author}{\bibfnamefont{P.}~\bibnamefont{Valente}},
  \bibinfo{author}{\bibfnamefont{H.}~\bibnamefont{Failache}},
  \bibinfo{author}{\bibfnamefont{M.}~\bibnamefont{Martinelli}},
  \bibnamefont{and}
  \bibinfo{author}{\bibfnamefont{P.}~\bibnamefont{Nussenzveig}},
  \bibinfo{journal}{Phys. Rev. A} \textbf{\bibinfo{volume}{77}},
  \bibinfo{pages}{013806} (\bibinfo{year}{2008}).

\bibitem[{\citenamefont{Mikhailov et~al.}(2009)\citenamefont{Mikhailov, Lezama,
  Noel, and Novikova}}]{Mikhailov09}
\bibinfo{author}{\bibfnamefont{E.~E.} \bibnamefont{Mikhailov}},
  \bibinfo{author}{\bibfnamefont{A.}~\bibnamefont{Lezama}},
  \bibinfo{author}{\bibfnamefont{T.~W.} \bibnamefont{Noel}}, \bibnamefont{and}
  \bibinfo{author}{\bibfnamefont{I.}~\bibnamefont{Novikova}},
  \bibinfo{journal}{Journal of Modern Optics} \textbf{\bibinfo{volume}{56}},
  \bibinfo{pages}{1985} (\bibinfo{year}{2009}).

\bibitem[{\citenamefont{Barreiro et~al.}(2011)\citenamefont{Barreiro, Valente,
  Failache, and Lezama}}]{Barreiro2011}
\bibinfo{author}{\bibfnamefont{S.}~\bibnamefont{Barreiro}},
  \bibinfo{author}{\bibfnamefont{P.}~\bibnamefont{Valente}},
  \bibinfo{author}{\bibfnamefont{H.}~\bibnamefont{Failache}}, \bibnamefont{and}
  \bibinfo{author}{\bibfnamefont{A.}~\bibnamefont{Lezama}},
  \bibinfo{journal}{Phys. Rev. A} \textbf{\bibinfo{volume}{84}},
  \bibinfo{pages}{033851} (\bibinfo{year}{2011}).

\bibitem[{\citenamefont{Labeyrie et~al.}(1999)\citenamefont{Labeyrie,
  de~Tomasi, Bernard, M{\"{u}}ller, Miniatura, and Kaiser}}]{Labeyrie1999}
\bibinfo{author}{\bibfnamefont{G.}~\bibnamefont{Labeyrie}},
  \bibinfo{author}{\bibfnamefont{F.}~\bibnamefont{de~Tomasi}},
  \bibinfo{author}{\bibfnamefont{J.-C.} \bibnamefont{Bernard}},
  \bibinfo{author}{\bibfnamefont{C.}~\bibnamefont{M{\"{u}}ller}},
  \bibinfo{author}{\bibfnamefont{C.}~\bibnamefont{Miniatura}},
  \bibnamefont{and} \bibinfo{author}{\bibfnamefont{R.}~\bibnamefont{Kaiser}},
  \bibinfo{journal}{Phys. Rev. Lett.} \textbf{\bibinfo{volume}{83}},
  \bibinfo{pages}{5266} (\bibinfo{year}{1999}).

\bibitem[{\citenamefont{Baudouin et~al.}(2013)\citenamefont{Baudouin,
  Mercadier, Guarrera, Guerin, and Kaiser}}]{baudouin2013cold}
\bibinfo{author}{\bibfnamefont{Q.}~\bibnamefont{Baudouin}},
  \bibinfo{author}{\bibfnamefont{N.}~\bibnamefont{Mercadier}},
  \bibinfo{author}{\bibfnamefont{V.}~\bibnamefont{Guarrera}},
  \bibinfo{author}{\bibfnamefont{W.}~\bibnamefont{Guerin}}, \bibnamefont{and}
  \bibinfo{author}{\bibfnamefont{R.}~\bibnamefont{Kaiser}},
  \bibinfo{journal}{Nature Physics} \textbf{\bibinfo{volume}{9}},
  \bibinfo{pages}{357} (\bibinfo{year}{2013}).

\bibitem[{\citenamefont{Guerin et~al.}(2009)\citenamefont{Guerin, Mercadier,
  Brivio, and Kaiser}}]{Guerin2009}
\bibinfo{author}{\bibfnamefont{W.}~\bibnamefont{Guerin}},
  \bibinfo{author}{\bibfnamefont{N.}~\bibnamefont{Mercadier}},
  \bibinfo{author}{\bibfnamefont{D.}~\bibnamefont{Brivio}}, \bibnamefont{and}
  \bibinfo{author}{\bibfnamefont{R.}~\bibnamefont{Kaiser}},
  \bibinfo{journal}{Opt. Express} \textbf{\bibinfo{volume}{17}},
  \bibinfo{pages}{11236} (\bibinfo{year}{2009}).

\bibitem[{\citenamefont{Labeyrie}(2008)}]{Labeyrie2008}
\bibinfo{author}{\bibfnamefont{G.}~\bibnamefont{Labeyrie}},
  \bibinfo{journal}{Mod. Phys. Lett. B} \textbf{\bibinfo{volume}{22}},
  \bibinfo{pages}{73} (\bibinfo{year}{2008}).

\bibitem[{\citenamefont{{Vernac, L.} et~al.}(2002)\citenamefont{{Vernac, L.},
  {Pinard, M.}, {Josse, V.}, and {Giacobino, E.}}}]{VERNAC02}
\bibinfo{author}{\bibnamefont{{Vernac, L.}}},
  \bibinfo{author}{\bibnamefont{{Pinard, M.}}},
  \bibinfo{author}{\bibnamefont{{Josse, V.}}}, \bibnamefont{and}
  \bibinfo{author}{\bibnamefont{{Giacobino, E.}}}, \bibinfo{journal}{Eur. Phys.
  J. D} \textbf{\bibinfo{volume}{18}}, \bibinfo{pages}{129}
  (\bibinfo{year}{2002}),
  \urlprefix\url{http://dx.doi.org/10.1140/e10053-002-0014-7}.

\bibitem[{\citenamefont{Josse et~al.}(2003)\citenamefont{Josse, Dantan, Vernac,
  Bramati, Pinard, and Giacobino}}]{JOSSE03}
\bibinfo{author}{\bibfnamefont{V.}~\bibnamefont{Josse}},
  \bibinfo{author}{\bibfnamefont{A.}~\bibnamefont{Dantan}},
  \bibinfo{author}{\bibfnamefont{L.}~\bibnamefont{Vernac}},
  \bibinfo{author}{\bibfnamefont{A.}~\bibnamefont{Bramati}},
  \bibinfo{author}{\bibfnamefont{M.}~\bibnamefont{Pinard}}, \bibnamefont{and}
  \bibinfo{author}{\bibfnamefont{E.}~\bibnamefont{Giacobino}},
  \bibinfo{journal}{Phys. Rev. Lett.} \textbf{\bibinfo{volume}{91}},
  \bibinfo{pages}{103601} (\bibinfo{year}{2003}),
  \urlprefix\url{http://link.aps.org/doi/10.1103/PhysRevLett.91.103601}.

\bibitem[{\citenamefont{Fabre}(1997)}]{FABRE97}
\bibinfo{author}{\bibfnamefont{C.}~\bibnamefont{Fabre}}, in
  \emph{\bibinfo{booktitle}{Quantum fluctuations}}
  (\bibinfo{publisher}{Elsevier Science B.V.}, \bibinfo{year}{1997}).

\bibitem[{\citenamefont{Mollow}(1969)}]{MOLLOW69}
\bibinfo{author}{\bibfnamefont{B.}~\bibnamefont{Mollow}},
  \bibinfo{journal}{Phys. Rev.} \textbf{\bibinfo{volume}{188}},
  \bibinfo{pages}{1969} (\bibinfo{year}{1969}).

\bibitem[{\citenamefont{Bimbard et~al.}(2014)\citenamefont{Bimbard, Boddeda,
  Vitrant, Grankin, Parigi, Stanojevic, Ourjoumtsev, and
  Grangier}}]{Bimbard_Homodyne_14}
\bibinfo{author}{\bibfnamefont{E.}~\bibnamefont{Bimbard}},
  \bibinfo{author}{\bibfnamefont{R.}~\bibnamefont{Boddeda}},
  \bibinfo{author}{\bibfnamefont{N.}~\bibnamefont{Vitrant}},
  \bibinfo{author}{\bibfnamefont{A.}~\bibnamefont{Grankin}},
  \bibinfo{author}{\bibfnamefont{V.}~\bibnamefont{Parigi}},
  \bibinfo{author}{\bibfnamefont{J.}~\bibnamefont{Stanojevic}},
  \bibinfo{author}{\bibfnamefont{A.}~\bibnamefont{Ourjoumtsev}},
  \bibnamefont{and} \bibinfo{author}{\bibfnamefont{P.}~\bibnamefont{Grangier}},
  \bibinfo{journal}{Phys. Rev. Lett.} \textbf{\bibinfo{volume}{112}},
  \bibinfo{pages}{033601} (\bibinfo{year}{2014}), \bibinfo{note}{and references
  therin}.

\bibitem[{\citenamefont{Dantan et~al.}(2005)\citenamefont{Dantan, Bramati, and
  Pinard}}]{DANTAN05}
\bibinfo{author}{\bibfnamefont{A.}~\bibnamefont{Dantan}},
  \bibinfo{author}{\bibfnamefont{A.}~\bibnamefont{Bramati}}, \bibnamefont{and}
  \bibinfo{author}{\bibfnamefont{M.}~\bibnamefont{Pinard}},
  \bibinfo{journal}{Phys. Rev. A} \textbf{\bibinfo{volume}{71}},
  \bibinfo{pages}{043801} (\bibinfo{year}{2005}).

\bibitem[{\citenamefont{Sargent~III et~al.}(1974)\citenamefont{Sargent~III,
  Scully, and Lamb~Jr.}}]{SARGENTBOOK74}
\bibinfo{author}{\bibfnamefont{M.}~\bibnamefont{Sargent~III}},
  \bibinfo{author}{\bibfnamefont{M.~O.} \bibnamefont{Scully}},
  \bibnamefont{and} \bibinfo{author}{\bibfnamefont{W.~E.}
  \bibnamefont{Lamb~Jr.}}, \emph{\bibinfo{title}{Laser Physics}}
  (\bibinfo{publisher}{Addison-Wesley Publishing Company},
  \bibinfo{address}{London}, \bibinfo{year}{1974}).

\bibitem[{\citenamefont{Cohen-Tannoudji
  et~al.}(1992)\citenamefont{Cohen-Tannoudji, Dupont-Roc, and
  Grynberg}}]{Cohen1992}
\bibinfo{author}{\bibfnamefont{C.}~\bibnamefont{Cohen-Tannoudji}},
  \bibinfo{author}{\bibfnamefont{J.}~\bibnamefont{Dupont-Roc}},
  \bibnamefont{and} \bibinfo{author}{\bibfnamefont{G.}~\bibnamefont{Grynberg}},
  \emph{\bibinfo{title}{Atom-Photon interactions, Basic processes and
  applications}} (\bibinfo{publisher}{John Wiley and Sons},
  \bibinfo{year}{1992}).

\bibitem[{Com()}]{Comment2}
\bibinfo{note}{Compared to what is reported in \citep{LEZAMA08}, we have not
  included in the atomic evolution the terms describing the arrival and
  departure of atoms to and from the system (\textit{i.e.}, $\gamma=0$). As a
  consequence, the steady state solution of the mean value of Eq. \ref{HL}
  requires the additional condition of a unity trace for the atomic density
  matrix.}

\bibitem[{\citenamefont{Collett et~al.}(1984)\citenamefont{Collett, Walls, and
  Zoller}}]{COLLETT84}
\bibinfo{author}{\bibfnamefont{M.~J.} \bibnamefont{Collett}},
  \bibinfo{author}{\bibfnamefont{D.~F.} \bibnamefont{Walls}}, \bibnamefont{and}
  \bibinfo{author}{\bibfnamefont{P.}~\bibnamefont{Zoller}},
  \bibinfo{journal}{Opt. Commun.} \textbf{\bibinfo{volume}{52}},
  \bibinfo{pages}{145} (\bibinfo{year}{1984}).

\bibitem[{\citenamefont{Heidmann and Reynaud}(1985)}]{HEIDMANN85}
\bibinfo{author}{\bibfnamefont{A.}~\bibnamefont{Heidmann}} \bibnamefont{and}
  \bibinfo{author}{\bibfnamefont{S.}~\bibnamefont{Reynaud}},
  \bibinfo{journal}{J. Phys. France} \textbf{\bibinfo{volume}{46}},
  \bibinfo{pages}{1937} (\bibinfo{year}{1985}).

\bibitem[{\citenamefont{Ho et~al.}(1987)\citenamefont{Ho, Kumar, and
  Shapiro}}]{HO87}
\bibinfo{author}{\bibfnamefont{S.-T.} \bibnamefont{Ho}},
  \bibinfo{author}{\bibfnamefont{P.}~\bibnamefont{Kumar}}, \bibnamefont{and}
  \bibinfo{author}{\bibfnamefont{J.}~\bibnamefont{Shapiro}},
  \bibinfo{journal}{Phys. Rev. A} \textbf{\bibinfo{volume}{35}},
  \bibinfo{pages}{3982} (\bibinfo{year}{1987}).

\end{thebibliography}

\end{document}